\documentclass[apl, twocolumn, amsmath,amssymb,amsfonts, superscriptaddress]{revtex4}
\usepackage[german,american,english]{babel}
\usepackage{graphicx}
\usepackage{graphics}
\usepackage{dcolumn}
\usepackage{bm}
\usepackage{amssymb}
\usepackage{amsmath}
\usepackage{amsfonts}
\usepackage{epsfig}
\newcommand {\ket} [1] {| #1 \rangle}
\newcommand {\bkt} [1] {\langle #1 \rangle}

\newcommand {\tbkt} [3] {\langle #1 | #2 | #3 \rangle}
\newcommand {\pd} [2] {\frac{\partial #1}{\partial #2}}

 \newcommand {\beq}{\begin{equation}}
\newcommand {\eeq}{\end{equation}}
\begin{document}
\title{Dephasing of Si singlet-triplet qubits due to charge and spin defects}
\author{Dimitrie Culcer}
\affiliation{ICQD, Hefei National Laboratory for Physical Sciences at the Microscale, University of Science and Technology of China, Hefei 230026, Anhui, China}
\affiliation{School of Physics, The University of New South Wales, Sydney 2052, Australia}
\author{Neil~M.~Zimmerman}
\affiliation{National Institute of Standards and Technology, Gaithersburg, Maryland 20899}
\begin{abstract}
We study the effect of charge and spin noise on singlet-triplet qubits in Si quantum dots. We set up a theoretical framework aimed at enabling experiment to efficiently identify the most deleterious defects, and complement it with the knowledge of defects gained in decades of industrial and academic work. We relate the dephasing rates $\Gamma_\phi$ due to various classes of defects to experimentally measurable parameters such as charge dipole moment, spin dipole moment and fluctuator switching times. We find that charge fluctuators are more efficient in causing dephasing than spin fluctuators. 
\end{abstract}
\date{\today}
\maketitle 

Quantum information processing is a powerful driving force spurring the development of quantum control of two-level systems in order to engineer and entangle quantum bits (qubits). The quest for scalable systems has led naturally to solid state quantum computing platforms. Among these, a substantial effort is underway researching Si spin quantum computing architectures, motivated by their compatibility with Si microelectronics and their long coherence times  \cite{Feher_PR59, Abe_PRB04, Tyryshkin_JPC06, Hanson_RMP07, Wang_SiQD_ST_Relax_PRB10, Raith_SiQD_1e_SpinRelax_PRB11}, thanks to the absence of piezoelectric electron-phonon coupling \cite{Prada_PRB08}, weak spin-orbit interaction \cite{Wilamowski_Si/SiGeQW_Rashba_PRB02, Tahan_PRB05} and the possibility of isotopic purification to remove the hyperfine interaction \cite{Witzel_AHF_PRB07}. Experimental breakthroughs have been reported in recent years in Si quantum dot (QD) and donor systems \cite{Zwanenburg_SiQmEl_RMP13, Morton_Si_QC_QmLim_Nat11}. The effort has focused mostly on single-spin qubits \cite{Loss_PRA98} and singlet-triplet qubits \cite{Petta_Science05}. 

The interface, substrate and gates are an integral part of the device architecture and influence its operation. A host of defects, which can be described as two-level fluctuators (TLFs), reside in these regions \cite{Sze} and cause qubit dephasing \cite{Sousa_BookChapter_09}, which is conventionally quantified by a dephasing rate $\Gamma_\phi \equiv 1/T_2^*$. The interaction of singlet-triplet qubits with the environment has been of intense interest of late \cite{Hu_PRL06, Culcer_APL09, RamonHu_DQD_Decoh_PRB10}. Singlet-triplet qubits exploit the charge and spin degrees of freedom simultaneously, relying on the exchange interaction and detuning for $\sigma_z$ rotations, and on an inhomogeneous magnetic field for $\sigma_x$ rotations. Consequently, the $\sigma_x$ gate is primarily affected by spin noise, while the $\sigma_z$ gate is primarily affected by charge noise. Spin noise is equivalent to a fluctuating inhomogeneous magnetic field, which couples directly to the qubit. The coupling of charge noise to the qubit depends on detuning, which is controllable. 

Previous work on noise-induced qubit decoherence has developed generic models of dephasing for single-spin and singlet-triplet qubits \cite{Hu_PRL06, Culcer_APL09, RamonHu_DQD_Decoh_PRB10}, with the result that the theoretical formulation of dephasing due common forms of noise, such as random telegraph and $1/f$ noise, is well understood. To date, however, theoretical models have not included realistic parameters for quantum dots in specific materials, and we are not aware of any work that has considered and compared the effects of coexisting charge and spin noise. Consequently, generic theoretical findings are not readily translated into information that can be of use to experiment. 

In light of the above observations, in this work we focus on the effect of noise from a realistic set of defects on qubit coherence. We analyse \textit{known} types of defects and identify the most deleterious ones in such a way as to be useful to experiment. Specifically, the present work aims to use knowledge garnered from decades of metal-oxide semiconductor (MOS) materials science and identify the most deleterious defects for the operation of singlet-triplet qubits: whether it is those that are electrically or magnetically active, those in the vicinity of the qubit, or those with the fastest switching times. We identify the main categories of TLFs and their associated dephasing times based on measurable parameters. Within the range of experimentally controllable parameters charge noise is considerably more effective than spin noise in inducing dephasing, which suggests future defect characterization should focus on charge noise.

\begin{figure}[tbp]
\includegraphics[width=\columnwidth]{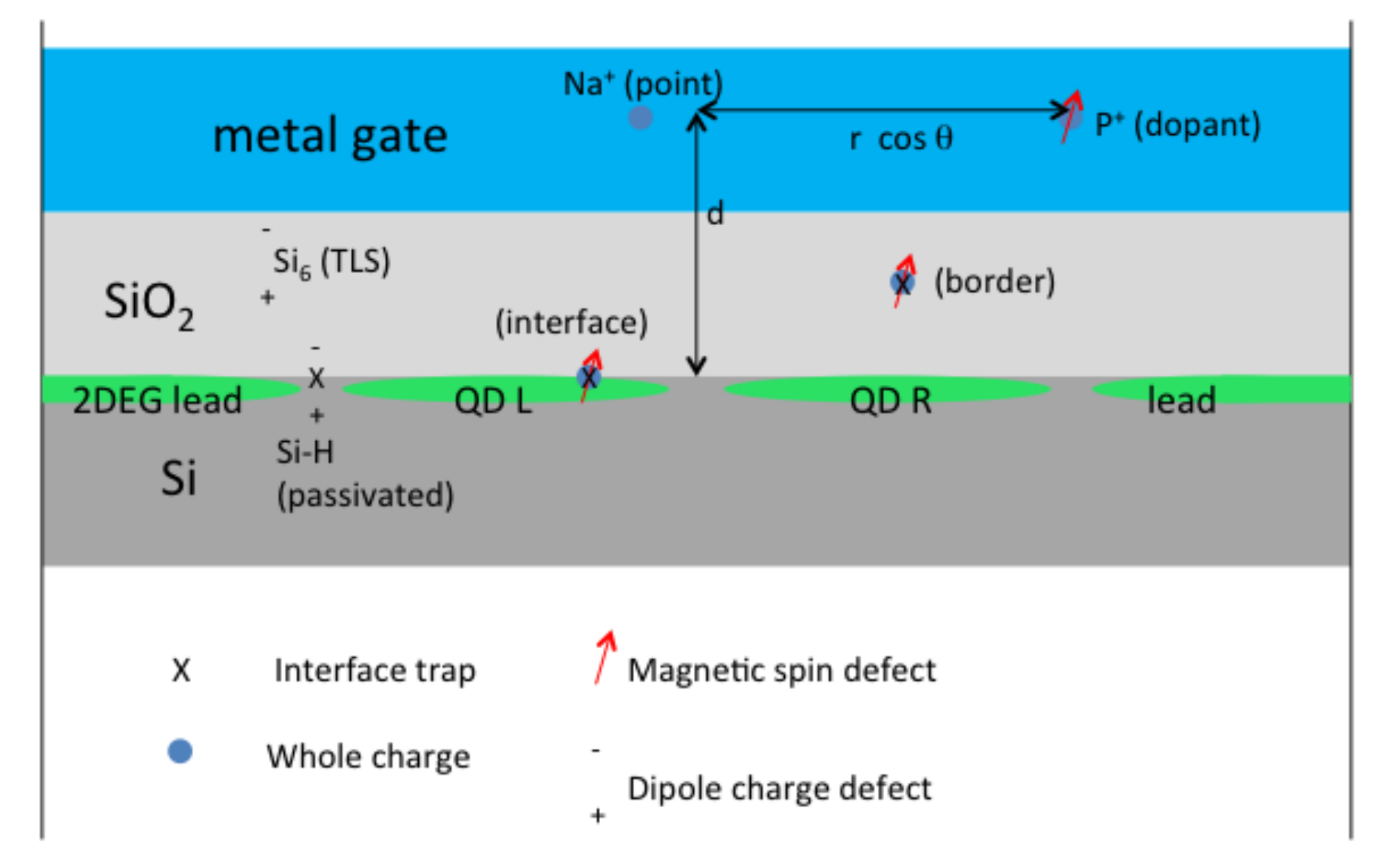}
\caption{\label{DS}
Sketch of defect locations with respect to the device and positions of defects projected onto the $xz$-plane.
}
\end{figure}

\begin{figure}[tbp]
\includegraphics[width=\columnwidth]{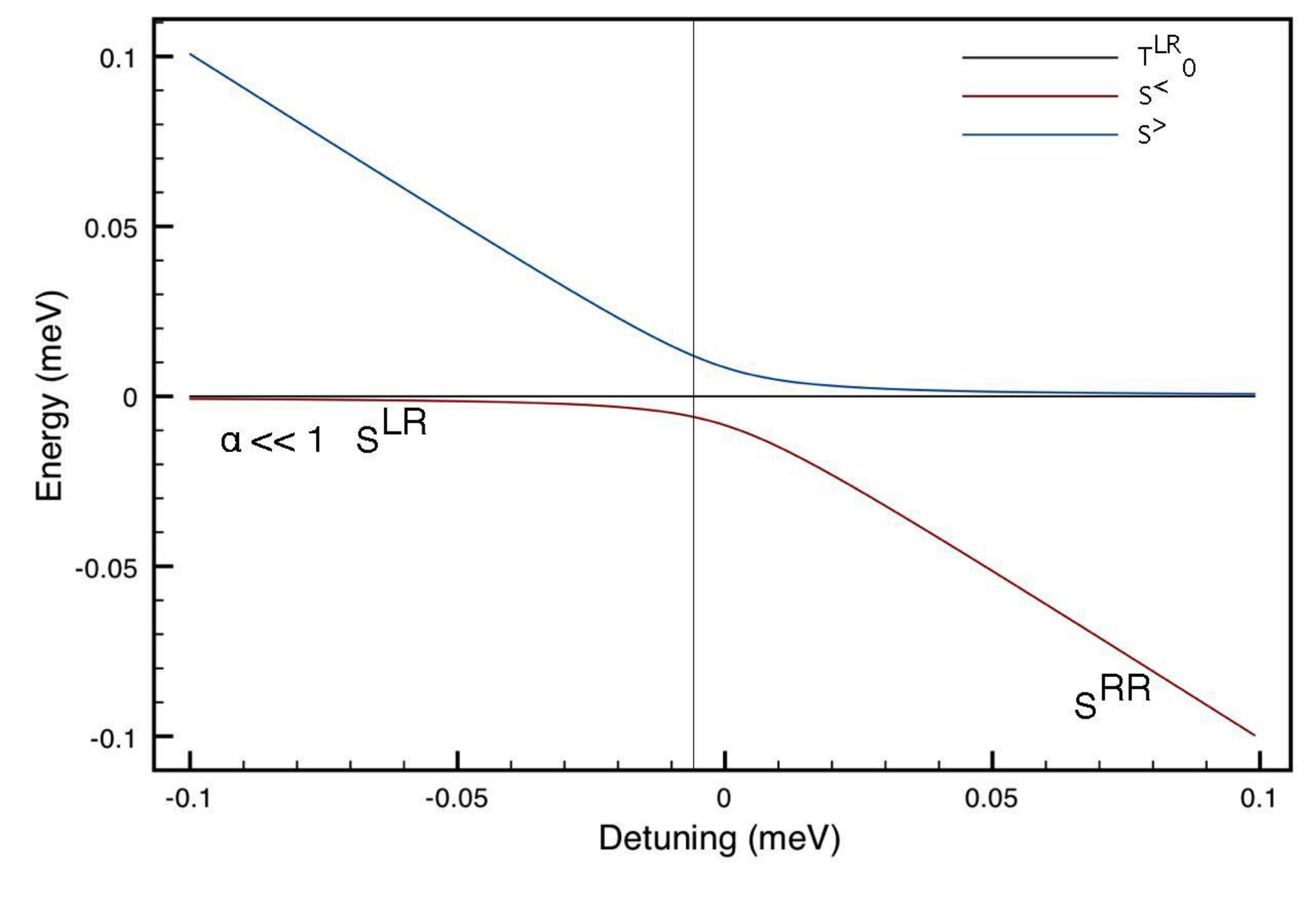}
\caption{\label{Explanation}
Operation of the singlet-triplet qubit. The system is initialized in the low-energy singlet $\ket{S^<}$, which, when $\delta > 0$, is predominantly $\ket{S^{RR}}$. As the detuning is swept into the far-detuned regime, $\ket{S^<}$ becomes predominantly $\ket{S^{LR}}$ when $|\delta| \gg t$ or $\alpha \ll 1$. This, together with the unpolarized triplet $\ket{T_0^{LR}}$, form the two qubit states (in the far-detuned regime). The high-energy singlet $\ket{S^>}$ is not involved in manipulation at all. The vertical line shows the (negative) minimum value of $t$.}
\end{figure}

We consider a double QD (DQD) with the left/right dots located at ${\bm R}_{L, R}$ respectively as in Fig.~\ref{DS}. We assume the valley-orbit coupling is sufficiently large as to allow us to consider only single-valley physics. The spatial parts of the one-electron wave functions centered at ${\bm R}_{L, R}$ are denoted by $\ket{L}$, $\ket{R}$ respectively. We use $(n, m)$ to refer to $n$ electrons on the left dot and $m$ electrons on the right dot. The (1,1) pure singlet and triplet states are
\begin{equation}
\arraycolsep 0.3 ex
\begin{array}{rl}
\displaystyle \ket{S^{LR}} = & \displaystyle (1/\sqrt{2}) \, \big( \ket{L^{(1)}R^{(2)}} + \ket{L^{(2)}R^{(1)}} \big) \, \ket{\chi_S} \\ [1ex] 
\displaystyle \ket{T^{LR}} = & \displaystyle (1/\sqrt{2}) \, \big( \ket{L^{(1)}R^{(2)}} - \ket{L^{(2)}R^{(1)}} \big) \, \ket{\chi_T},
\end{array}
\end{equation}
where $(1)$ and $(2)$ label the electrons and $\ket{\chi_S}$ and $\ket{\chi_T}$ are spin singlet and triplet states \cite{Supplement}. The (0,2) singlet, $\ket{S^{RR}} = \displaystyle \ket{R^{(1)}R^{(2)}} \, \ket{\chi_S}$, is involved in qubit manipulation, which is described in the caption of Fig.~\ref{Explanation}.

A Zeeman field is assumed to split off the polarized triplets. The two-electron Hamiltonian restricted to the three states $\{\ket{S^{LR}}, \ket{S^{RR}}, \ket{T_0^{LR}} \}$ has the form
\begin{equation}
H_{2e} = \begin{pmatrix}
j & t \sqrt{2} & 0 \cr
t \sqrt{2} & - \delta & 0 \cr
0 & 0 & - j \cr
\end{pmatrix}.
\end{equation}
Here $t$ is the interdot tunneling, $\delta$ is the detuning, and the exchange integral $j = \displaystyle \tbkt{L^{(1)}  R^{(2)}}{V_{ee}}{L^{(2)}  R^{(1)}}$.
The potential $V_{ee} = \displaystyle \frac{e^2}{4\pi\varepsilon_0\varepsilon_r |{\bm r}_1 - {\bm r}_2|}$ is the Coulomb interaction between two electrons located at ${\bm r}_1$ and ${\bm r}_2$, with $\varepsilon_0$ the permittivity of free space and $\varepsilon_r$ the relative permittivity. The eigenvalues of $H_{2e}$ are $-j$ and
\begin{equation}
\begin{array}{rl}
\displaystyle \varepsilon^< = \frac{j - \delta - \sqrt{(j + \delta)^2 + 8t^2} }{2} \\ [3ex]
\displaystyle \varepsilon^> = \frac{j - \delta + \sqrt{(j + \delta)^2 + 8t^2} }{2}.
\end{array}
\end{equation}
The eigenstates are the unpolarized triplet $\ket{T_0^{LR}}$ and
\begin{equation}
\begin{array}{rl}
\displaystyle \ket{S^<} = & \displaystyle \frac{\varepsilon^<}{\sqrt{\varepsilon^{<2} + 2t^2}} \, \bigg( \frac{t \sqrt{2}}{\varepsilon^<} \, \ket{S^{LR}} + \ket{S^{RR}} \bigg) \, \ket{\chi_S} \\ [3ex]
\displaystyle \ket{S^>} = & \displaystyle \frac{\varepsilon^>}{\sqrt{\varepsilon^{>2} + 2t^2}} \, \bigg( \frac{t \sqrt{2}}{\varepsilon^>} \, \ket{S^{LR}} + \ket{S^{RR}} \bigg) \, \ket{\chi_S}.
\end{array}
\end{equation} 
By matrix elements in the \textit{qubit subspace} we understand matrix elements involving the two eigenstates $\ket{S^<}$ and $\ket{T^{LR}_0}$. The state $\ket{S^>}$ is of no interest in this work. In the far detuned regime the detuning $\delta$ is negative, $\delta \ll 0$, $|\delta| \gg t$, and $t \gg \varepsilon^<$. In that regime the qubit states are the pure unpolarized triplet $\ket{T^{LR}_0}$ and the singlet $\ket{S^<} \approx \ket{S^{LR}} - \alpha \, \ket{S^{RR}}$, where the admixture ratio $\alpha^2 = 2 \, (t/|\delta|)^2 \ll 1$. In the \textit{qubit subspace}, in the basis $\{\ket{S^<}, \ket{T^{LR}_0} \}$, the effective two-electron Hamiltonian is $H_{Qbt} = (1/2) \, (\varepsilon^< + j) \, \sigma_z$. The vector ${\bm \sigma} \equiv \{\sigma_x, \sigma_y, \sigma_z\}$ of Pauli spin matrices here refers to the qubit subspace. 

Spin noise couples directly to the qubit and yields fluctuations in $\hat{\bm x}$-rotations. Charge fluctuations couple to a purely spin qubit through $\alpha$, since charge noise modulates the energy splitting $\delta$ between the dots and thus $\varepsilon^<$ through a fluctuation $\Delta\varepsilon^<$, yielding fluctuations in $\hat{\bm z}$-rotations. Specifically, the dephasing rates $\Gamma_\phi^{wh}$, $\Gamma_\phi^{dip}$ due to whole and dipole charge defects are $\propto \alpha^4$. Given that $\alpha \ll 1$, one might expect charge noise effects to be small, in particular since it is commonly assumed that $\delta$ can be made arbitrarily large. Recent work, however, shows that charge noise is indeed important \cite{Dial_PRL13}. Moreover, given that the coupling of the charge to electrical noise (governed by the quantity $e^2/4\pi\varepsilon_0\varepsilon_r$) is much stronger than the coupling of the spin to magnetic noise (governed by the Bohr magneton), the relative effects of charge and spin noise in singlet-triplet qubits are not obvious \textit{a priori}.

The total Hamiltonian
\begin{equation}\label{H}
H = H_{Qbt} + \frac{1}{2} \, {\bm \sigma} \cdot {\bm V}(t),
\end{equation}
where ${\bm V}(t)$ stems from noise. For a charge defect ${\bm \sigma} \cdot {\bm V}(t) = V \, \sigma_z \, (-1)^{N(t)}$, and the switching of ${\bm V}(t)$ is quantified by a Poisson random variable $N(t) = 0, 1$ with an average switching time $\tau$. We define $h (t) = \frac{1}{\hbar} \int_0^t V(t') \, dt' = \frac{V}{\hbar} \int_0^{t'} dt' \, (-1)^{N(t')}$. To study dephasing in the spin expectation value $S_x$, we determine the time dependence 
\begin{equation}
S_x (t) = {\rm tr} \, \sigma_x \rho (t) = S_{0x} \, \cos h(t),
\end{equation}
where $S_{0x}$ is the initial spin. To determine $\Gamma_\phi$ we average $\cos h(t)$ over the realizations of $h(t)$, denoted by $\bkt{\cos h(t)}$. In singlet-triplet qubits in the (1,1) regime, $\varepsilon^<$ is of the order of neV, corresponding to several microseconds \cite{Petta_Science05}, thus $V \approx \Delta\varepsilon^< \approx$ neV \cite{Culcer_APL09}. With the upper bound for $\tau$ at 1 $\mu$s \footnote{We take 1 $\mu$s as a cutoff for the fluctuator switching time, since experimentally the effect of fluctuators with switching times longer than $\approx 1$ $\mu$s can be eliminated by means of dynamical decoupling. For short switching time, $V\tau \ll \hbar$, the time dynamics of $h(t)$ are described by a random walk in time, in which the spread in $\cos h(t)$ leads to motional narrowing.}, $V^2 \ll (\hbar/\tau)^2$ and \cite{Sousa_PRB03} 
\begin{equation}\label{avg}
\bkt{\cos h(t)} = \displaystyle e^{-\frac{t}{\tau}} \, \bigg( \frac{\sinh \Omega t }{\Omega \tau} + \cosh \Omega t \bigg),
\end{equation}
where $\Omega = \displaystyle \sqrt{1/\tau^2 - V^2/\hbar^2}$. Expanding $\Omega$ in $V^2\tau^2/\hbar^2$, ignoring terms $\propto V^2\tau^2$ in the denominator in Eq.\ (\ref{avg}), $\bkt{\cos h(t)} \approx e^{- (\frac{V^2\tau}{2\hbar^2}) t}$. The dephasing rate for one TLF
\begin{equation}\label{T2*}
\Gamma_\phi = \frac{V^2\tau}{2\hbar^2},
\end{equation}
whereupon $S_i (t) = S_i(0) e^{- \Gamma_\phi t}$, true for both $\sigma_x$ and $\sigma_z$ noise. In this motional narrowing regime, defects with the longest switching times are the most deleterious, and defects with faster switching times are less important.

\begin{figure}[tbp]
\includegraphics[width=\columnwidth]{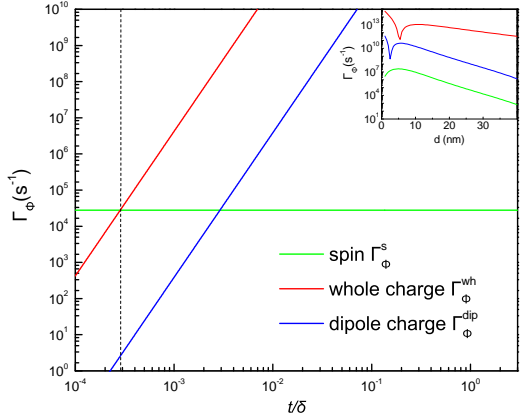}
\caption{\label{Gamma}
Dephasing rates $\Gamma_\phi$ as a function of $t/|\delta|$ for a single defect, $X_0 = 20$ nm, $r = 30$ nm, $d = 3$ nm, and $\theta = 0$. The vertical line denotes the crossover from charge-defect dominated to spin-defect dominated dephasing at $t/|\delta| \approx 3 \times 10^{-4}$. The inset shows $\Gamma_\phi$ as a function of $d$ for $t/|\delta| = 0.06$ (note that the three lines never cross).}
\end{figure}

Fig.~\ref{DS} is a sketch of a typical environment. The most important known classes of defects include interface traps, tunneling two-level systems (TLS), dislocations, grain boundaries, dopants, mobile ions and magnetic dipoles \cite{Sze, Fleetwood}. Dislocations and grain boundaries are static and do not contribute to charge fluctuations. Defects are broadly grouped into whole charge fluctuators, dipole charge fluctuators and spin fluctuators, and their associated quantities are denoted by $wh$, $dip$ and $s$ respectively. Quantities that depend explicitly on defect class (see Table I) include the height $d$, whole charge change $Q$ and magnetic dipole change ${\bm \mu}$, the charge dipole length $l$, the switching time $\tau$, and typical areal density $n$. On the other hand  $t$, $\delta$, and $X_0$ \cite{Supplement} depend solely on the qubit.

We analyze dephasing due to each type using Eq.\ (\ref{T2*}). Spin defects are quantified by their (dimensionless) spin ${\bm s}_d$. For a single defect \cite{Supplement}
\begin{equation}\label{Gammaphis}
\arraycolsep 0.3ex
\begin{array}{rl}
\displaystyle \Gamma_\phi^s = \frac{\tau \, (\mu_0 g_e^2\mu_B^2)^2}{32\pi^2\hbar^2} \, \bigg\{ & \displaystyle \bigg[\frac{3d ({\bm s}_d\cdot\hat{\bm R}_{L}) }{R_{L}^4} - \frac{s_{d, z}}{R_{L}^3}\bigg]  - (L \leftrightarrow R) \bigg\}^2.
\end{array}
\end{equation}
For a uniform areal density $n_s$ of identical spin defects, the dephasing rate $\Gamma_\phi^{n, s} = n_s \int dr \, r \int d\theta \, \Gamma_\phi^s$.

Charge defects are screened by a nearby two-dimensional electron gas (2DEG), which we account for in the random phase approximation, with the (constant) Si Thomas-Fermi wave vector $q_{TF} \approx 0.6$ nm$^{-1}$. We do not include screening by nearby gates. $\Gamma_\phi^{wh/dip}$ for a single charge defect and $\Gamma_\phi^{n, {wh/dip}}$ for a uniform charge defect density, found analogously to spin defects, are given explicitly in the Supplement \cite{Supplement}. For charge defects the Friedel oscillation part of the potential makes a negligible contribution to $\Gamma_\phi$ as compared to its regular part. The orientation of the charge/spin dipole moment with respect to the qubit also makes little difference for $\Gamma_\phi$.

\begin{table}[tbp]
\label{DefectTable}
\caption{Defect classes and associated relative dephasing rates normalized to a base rate of $10^{15}$ s$^{-1}$. Unless specifically noted, attribute values are based on Refs.~\onlinecite{Sze} and \onlinecite{Fleetwood}. Here $d$ refers to the vertical separation between the defect and the QD. Referring to Fig.~4 in the Supplement \cite{Supplement}, we assume $X_0 = $ 20 nm, the Thomas-Fermi wave vector $q_{TF} = 0.6$ nm$^{-1}$, and a reference $\tau = 1\, \mu$s.}

 $ \arraycolsep 0.15em
   \begin{array}{c@{\hspace{2em}}ccccccccc} \hline\hline

Attribute & TLS & Intf. & Pass. & Border & Dopant & Pt. \\ \hline

n\ (cm^{-2}) & 10^{11}$ \cite{Zimmermann_PRL81} $& 10^{11} & 10^{12} & 10^{11} & 10^9 & 10^{12} $\cite{Jang_JES82}$ \\ [1ex]

l\ (nm) & 0.093 $\cite{Reinisch_JPCB06}$ & 0 & 0.15 $\cite{Biswas_PRL99}$ & 0 & 0 & 0 \\ [1ex]

Q\ (e) & 0 & 1 & 0 & 1 & 1 & 1 \\ [1ex]

d\ (nm) & 10 & 3 & 3 & 5& 10 & 10 \\ [1ex]

\mu/(g\mu_B) & 0 & 2 & 0 & 2 & 2 & 0 \\ [1ex]

\Gamma_\phi^{n,{wh}} & 0 & 1 & 0 & 0.5 & 10^{-3} & 1 \\ [1ex] 

\Gamma_\phi^{n,{dip}} & 5\times 10^{-4} & 0 & 10^{-2} & 0 & 0 & 0 \\ [1ex]

\Gamma_\phi^{n,s} & 0 & 10^{-9} & 0 & 5 \times 10^{-11} & 10^{-14} & 0

 \\ \hline\hline
  \end{array}$
\end{table}

It is beyond the scope of this work to determine the overall magnitudes of switching times for individual classes of defects. However, we can gain insight on their relative magnitudes by considering a typical defect having a whole charge, a charge dipole moment and a spin dipole moment. If we assume the in-plane separation of the defect from the QD to be much larger than their vertical separation, we find $\Gamma^{dip}_\phi/\Gamma^{wh}_\phi \approx 10^{-3}$ and $\Gamma^{s}_\phi/\Gamma^{wh}_\phi \approx 10^{-10}$, indicating that charge noise is much more efficient at inducing dephasing than spin noise, and that whole charges are more efficient than charge dipoles. Fig.~\ref{Gamma} illustrates this by plotting the single-defect spin, whole, and dipole charge dephasing rates as a function of $t/|\delta|$ and $d$. For a set $d=3$ nm, the charge dephasing rates exceed the spin dephasing rate beyond very small values of $t/|\delta|$. Experimentally, the maximum $\delta$ is set by a physical energy scale, here the valley splitting of $0.1$ meV, while the minimum $t$ is set by the gate operation speed of $1/(10^4 \, T_2^*)$. Estimating the maximum $T_2^*$ as $10^{-5}$ s yields $t/|\delta| \ge 0.06$ so $\alpha^2 \ge 0.007$. Thus, at this fixed $d$, in any realistic experimental setting, the dephasing rates due to charge defects will be much larger than those due to spin defects. 

The inset of Fig.~\ref{Gamma} plots the dependence of the dephasing rates on $d$. Since the lines never cross, Fig.~\ref{Gamma} tells us that our qualitative conclusions in the previous paragraph are true for the entire range of possible values of $d$ shown. Fig.~\ref{Gamma} allows us to suggest ameliorating the coherence times by e.g. increasing the thickness of the layer so as to move defects further away, and by passivation of the most deleterious defects: interface, border and mobile point defects.

Sample dephasing rates for uniform defect densities have been calculated in the Table \ref{DefectTable} using typical defect parameter values. All defects have been assigned a switching time $\tau = 1$ $\mu$s. In this work we do not consider a distribution of switching times $\tau$, but rather focus on an incoherent array of fluctuators, each giving rise to a random walk in time, all with the \textit{same} $\tau$. This allows us to average $\Gamma_\phi$, rather than $e^{-\Gamma_\phi t}$, over defect locations. The Table \ref{DefectTable} upholds our findings that charge defects are much more efficient in causing dephasing than spin defects. 

Conventional wisdom holds that, since magnetic defects couple to the spin degree of freedom, their effect on dephasing of spin qubits should be stronger. Our analysis of both individual defects and uniform defect densities shows this to be incorrect. Charge defects, including interface, border and mobile point defects, are most deleterious for singlet-triplet qubits. The effect of charge defects on dephasing can be reduced by going into the far-detuned regime, since the dephasing rate is $\propto (t/|\delta|)^4$. At the same time, the exchange coupling $\propto (t^2/|\delta|)$, thus in the far detuned regime the effect of charge noise is weaker, but with the cost that the exchange gate is correspondingly slower.

In summary we have constructed a picture of the potential landscape seen by singlet-triplet qubits in Si and conclude that charge defects are much more effective in inducing dephasing than spin defects. In the process of analyzing a specific architecture and material system, we laid out a general framework enabling this analysis to be extended to different types of devices, so that the most deleterious defects can be identified in other types of charge and spin qubits.

We are grateful to Vanita Srinivasa (NIST), Emily Townsend (NIST), S.~Das Sarma, and Qian Han for enlightening discussions. This work is supported by the National Natural Science Foundation of China under grant number 91021019. 


\begin{widetext}

\section{Supplement to: Dephasing of Si singlet-triplet qubits due to charge and spin defects}

\end{widetext}

\subsection{Spin wave functions}

The singlet and triplet spin wave functions are 
\begin{equation}
\arraycolsep 0.3 ex
\begin{array}{rl}
\displaystyle \ket{\chi_S} = & \displaystyle \frac{1}{\sqrt{2}} \, ( \ket{\uparrow^{(1)}\downarrow^{(2)}} - \ket{\downarrow^{(1)}\uparrow^{(2)}}) \\ [3ex]
\displaystyle \ket{\chi_{T, \uparrow \uparrow}} = & \displaystyle \ket{\uparrow^{(1)}\uparrow^{(2)}} \\ [3ex]
\displaystyle \ket{\chi_{T, \downarrow \downarrow}} = & \displaystyle \ket{\downarrow^{(1)}\downarrow^{(2)}} \\ [3ex]
\displaystyle \ket{\chi_{T, 0}} = & \displaystyle \frac{1}{\sqrt{2}} \, ( \ket{\uparrow^{(1)}\downarrow^{(2)}} + \ket{\downarrow^{(1)} \uparrow^{(2)}}). 
\end{array}
\end{equation}
We use $\ket{\chi_T}$ generically for any of the three triplet wave functions. 

\subsection{Qubit time evolution}

We work in a \textit{rotating} frame of reference, in which the effect of the rotation $(1/2) \, (\varepsilon^< + j) \, \sigma_z$ has been eliminated. We subsequently follow the time evolution of the spin, in the rotating frame of reference, in a free induction decay experiment. The time evolution of the density matrix is given by 
\begin{equation}\label{rhot}
\begin{array}{rl}
\displaystyle \rho (t) = & \displaystyle e^{-i\int_0^t H(t') \, dt'} \rho(0) \, e^{i\int_0^t H(t') \, dt'}.
\end{array}
\end{equation}
We need to calculate the time evolution of the components $i$ of the spin, $S_i = {\rm tr} \, \sigma_i \rho$, then particularize to specific initial conditions. Using the summation convention we obtain
\begin{equation}\label{sigmat}
S_i (t) = \frac{S_{0j}}{2}\, {\rm tr} \, \sigma_i \, e^{-\frac{i}{\hbar}\int_0^t H(t') \, dt'} \sigma_j \, e^{\frac{i}{\hbar}\int_0^t H(t') \, dt'},
\end{equation}
where $S_{0j}$ is the initial value of $S_j$. The time evolution operator can be written as
\begin{equation}\label{exp}
\begin{array}{rl}
\displaystyle e^{-\frac{i}{\hbar} \int_0^t H(t') \, dt'} = & \displaystyle \cos\bigg[\frac{h(t)}{2}\bigg] - i {\bm \sigma} \cdot \hat{\bm h}(t) \, \sin\bigg[\frac{h(t)}{2}\bigg], 
\end{array}
\end{equation}
and therefore, after some algebra, Eq.\ (\ref{sigmat}) becomes
\begin{equation}
\begin{array}{rl}
\displaystyle S_i (t) = & \displaystyle S_{0i} \, \cos h(t) - \epsilon_{ijk} S_{0j} \hat{h}_k \sin h(t) \\ [1ex]
+ & \displaystyle \hat{h}_i (t) [\hat{\bm h}(t) \cdot {\bm S}_0] \, [1 - \cos h(t)],
\end{array}
\end{equation}
where $\hat{\bm h}$ denotes the unit vector ${\bm h}/|{\bm h}|$. 

\begin{figure}[tbp]
\includegraphics[width=0.9\columnwidth, height = 200pt]{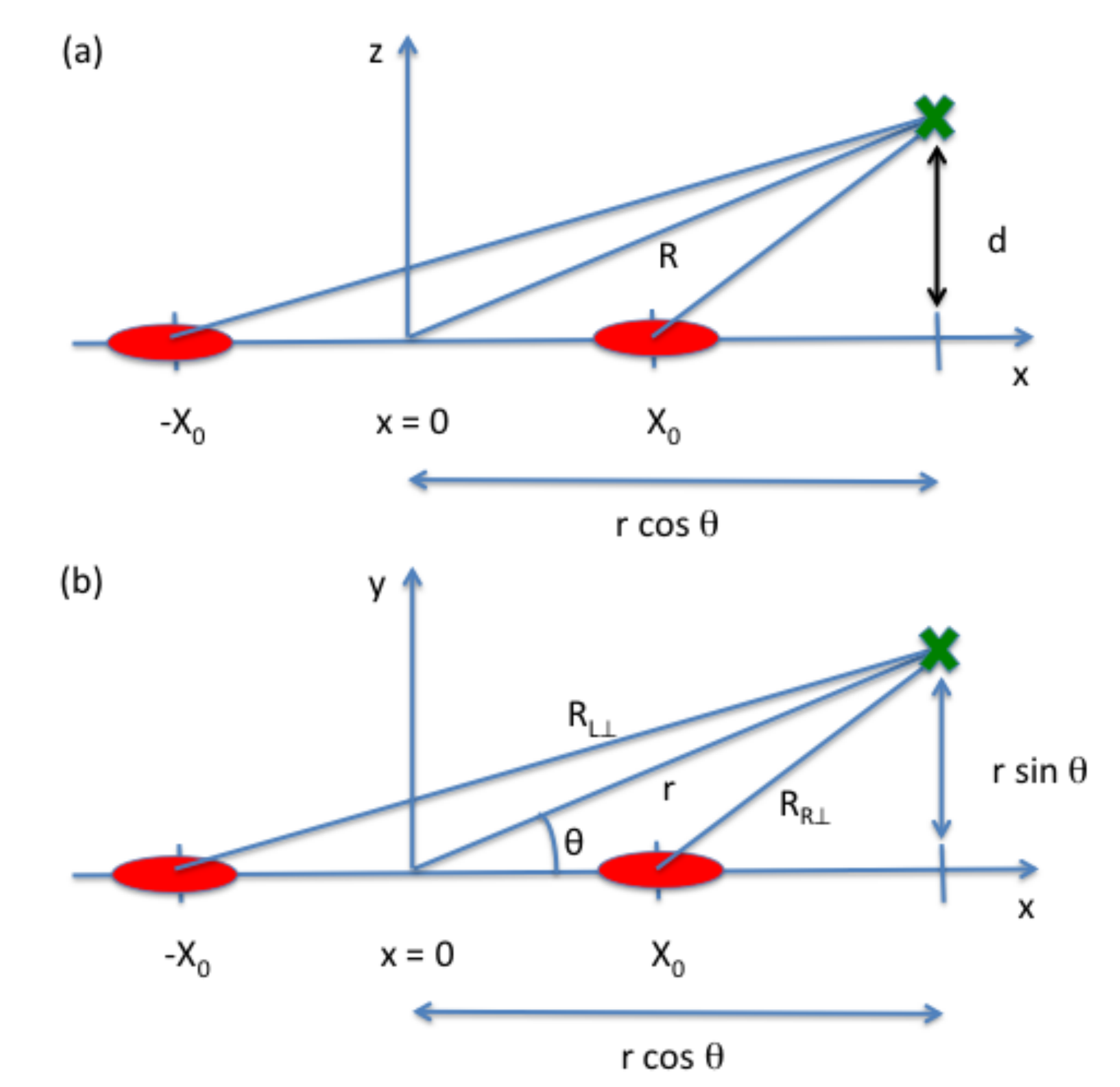}
\caption{\label{X0} Defect geometry and notation. The location of the defect is shown as a green cross. The double QD lies in the $xy$-plane. (a) Projection of defect position onto $xz$-plane; (b) Projection of defect position onto $xy$-plane.
}
\end{figure}

\subsection{Effective spin defect potentials} 
 
Using Fig.\ \ref{X0}, 
\begin{equation}\label{}
\arraycolsep 0.3 ex
\begin{array}{rl}
\displaystyle {\bm R}_L = & \displaystyle ({\bm R}_{L\perp}, d) \\ [3ex]
\displaystyle {\bm R}_R = & \displaystyle ({\bm R}_{R\perp}, d) \\ [3ex]
\displaystyle {\bm R}_{L\perp}^2 = & \displaystyle (r \cos \theta + X_0)^2 + r^2\sin^2\theta \\ [3ex]
\displaystyle {\bm R}_{R\perp}^2 = & \displaystyle (r \cos \theta - X_0)^2 + r^2\sin^2\theta \\ [3ex]
\end{array}
\end{equation}

The magnetic dipole corresponding to one spin ${\bm s}$ is ${\bm m} = - g_e \mu_B{\bm s}$. The gyromagnetic ratio is $\gamma_e = g_e\mu_B/\hbar$. The interaction between a spin in the dot and one in the defect is 
\begin{equation}
\begin{array}{rl}
\displaystyle H_{B, dip} = & \displaystyle - \frac{\mu_0 g_e^2\mu_B^2}{8\pi \hbar R_{dD}^3} \, {\bm \sigma} \cdot [3({\bm s}_d\cdot\hat{\bm R}_{dD}) \, \hat{\bm R}_{dD} - {\bm s}_d] \\ [3ex]
= & \displaystyle \frac{1}{2} \, \mu_B \, {\bm \sigma} \cdot {\bm B}_{dip}.
\end{array}
\end{equation}
We divide the effective magnetic field ${\bm B}_{dip}$ into a total part and a difference part between the dots: ${\bm B}_{dip, tot} = {\bm B}_{dip, L} + {\bm B}_{dip, R}$ and $\Delta{\bm B}_{dip} = {\bm B}_{dip, L} - {\bm B}_{dip, R}$. The qubit states are $\ket{S^{LR}}$ and $\ket{T_0^{LR}}$. The total field ${\bm B}_{dip, tot}$ mixes the triplets among themselves. The $\hat{\bm z}$-component of the difference $\Delta{\bm B}_{dip}$ mixes the admixture singlet with the unpolarized triplet. The $\hat{\bm x}$ and $\hat{\bm y}$ components of $\Delta{\bm B}_{dip}$ mix the admixture singlet with the polarized triplets. The polarized triplets are split by a lab magnetic field and are far away in energy. Therefore in principle we only care about the $\hat{\bm z}$-component of the magnetic field $\Delta{\bm B}_{dip}$, which mixes the admixture singlet and unpolarized triplet. The Hamiltonian due to this inhomogeneous magnetic field in the qubit subspace gives rise to matrix elements in the two-electron Hamiltonian
\begin{equation}\label{Bdip}
\arraycolsep 0.3 ex
\begin{array}{rl}
\displaystyle H_{B, ST} = & \displaystyle - \frac{1}{2} \, \mu_B \begin{pmatrix}
0 & \Delta B^{dip}_z \cr
\Delta B^{dip}_z & 0
\end{pmatrix}
\end{array}
\end{equation}
As a result, a fluctuation in this (real) magnetic field constitutes a fluctuating effective magnetic field $\parallel \hat{\bm x}$ in the qubit subspace. Unlike charge noise, $H_{B, ST}$ has nothing to do with tunneling.

\begin{widetext}

\subsubsection{Dephasing rate for spin defects}

For an areal density $n_s$ of spin defects with homogeneous switching time $\tau$
\begin{equation}\label{}
\arraycolsep 0.3 ex
\begin{array}{rl}
\displaystyle \Gamma_\phi^{n_s} = \frac{\tau n_s}{2\hbar^2} \, \bigg(\frac{\mu_0 g_e^2\mu_B^2}{4\pi}\bigg)^2 \, \int_0^\infty dr \, r \int_0^{2\pi} d\theta
& \displaystyle \bigg\{ \frac{1}{R_{L}^3} \, \bigg[\frac{3d}{R_{L}} \, ({\bm s}_d\cdot\hat{\bm R}_{L}) \, - s_{d, z}\bigg] - \frac{1}{R_{R}^3} \, \bigg[\frac{3d}{R_{R}} \, ({\bm s}_d\cdot\hat{\bm R}_{R}) \, - s_{d, z}\bigg]\bigg\}^2.
\end{array}
\end{equation}
We study the two extreme cases, first when ${\bm s}_d \parallel \hat{\bm x}$ (in-plane) and second ${\bm s}_d \parallel \hat{\bm z}$ (out-of-plane). For ${\bm s}_d \parallel \hat{\bm x}$, we define first the dimensionless variables $\tilde{r} = r/X_0$ and $\tilde{d} = d/X_0$, which yields $ \Gamma_\phi^{n_s} = \frac{9d^2 s_d^2 \tau n_s}{2\hbar^2 X_0^6} \, (\frac{\mu_0 g_e^2\mu_B^2}{4\pi})^2 I^s_x (\tilde{d}) $, with 
\begin{equation}\label{}
\arraycolsep 0.3 ex
\begin{array}{rl}
\displaystyle I^s_x = & \displaystyle \int_0^\infty d\tilde{r} \, \tilde{r} \int_0^{2\pi} d\theta \bigg[ \frac{\tilde{r} \cos\theta + 1}{[(\tilde{r} \cos \theta + 1)^2 + \tilde{r}^2\sin^2\theta + \tilde{d}^2]^{5/2}} - \frac{\tilde{r}\cos\theta - 1}{[(\tilde{r} \cos \theta - 1)^2 + \tilde{r}^2\sin^2\theta + \tilde{d}^2]^{5/2}} \bigg]^2.
\end{array}
\end{equation}
For ${\bm s}_d \parallel \hat{\bm z}$ we obtain $\Gamma_\phi^{n_s} = \frac{s_d^2 \tau n_s}{2\hbar^2 X_0^4} \, (\frac{\mu_0 g_e^2\mu_B^2}{4\pi})^2 \, I^s_z (\tilde{d}) $, where
\begin{equation}\label{}
\arraycolsep 0.3 ex
\begin{array}{rl}
\displaystyle I^s_z = & \displaystyle \int_0^\infty d\tilde{r} \, \tilde{r} \int_0^{2\pi} d\theta \bigg\{\bigg(\frac{3\tilde{d}^2}{[(\tilde{r} \cos \theta + 1)^2 + \tilde{r}^2\sin^2\theta + \tilde{d}^2]^{5/2}} -  \frac{1}{[(\tilde{r} \cos \theta + 1)^2 + \tilde{r}^2\sin^2\theta + \tilde{d}^2]^{3/2}} \bigg) \\ [3ex]
- & \displaystyle \bigg(\frac{3\tilde{d}^2}{[(\tilde{r} \cos \theta - 1)^2 + \tilde{r}^2\sin^2\theta + \tilde{d}^2]^{5/2}} - \frac{1}{[(\tilde{r} \cos \theta - 1)^2 + \tilde{r}^2\sin^2\theta + \tilde{d}^2]^{3/2}} \bigg) \bigg\}^2.
\end{array}
\end{equation}

\subsubsection{Dephasing rate for whole charge defects}

We use $V_{wh} = \alpha^2 \, (U_{scr}^L - U_{scr}^R)$. To describe a screened Coulomb potential we use its Fourier transform, 
\begin{equation}
\arraycolsep 0.3 ex
\begin{array}{rl}
\displaystyle U_{scr}(r) = & \displaystyle \int \frac{d^2 q}{(2\pi)^2} \, e^{-i {\bm q}\cdot{\bm r}} U_{scr}(q) \equiv \int \frac{d^2 q}{(2\pi)^2} \, U_{scr}(q) \,  \cos{\bm q}\cdot{\bm r} \\ [3ex]
\displaystyle U_{scr}(q) = & \displaystyle \frac{e^2}{2 \epsilon_0 \epsilon_r} \, e^{-qd} \bigg[\frac{\Theta(2k_F - q)}{q + q_{TF}} + \frac{\Theta(q - 2k_F)}{q + q_{TF} \, \big\{ 1 - \big[ 1 - \big( \frac{2k_F}{q} \big)^2 \big]^{1/2}  \big\} } \bigg],
\end{array}
\end{equation}
where $\Theta$ is the Heaviside function. For $r > 2X_0$
\begin{equation}\label{Uscrlarge}
U_{scr} (r) = \frac{e^2 q_{TF}}{4\pi \epsilon_0 \epsilon_r} \, \bigg[\frac{1 + q_{TF}d}{(q_{TF}r)^3} - e^{-2k_Fd} \frac{\sin2k_Fr}{(2k_Fr)^2}\bigg].
\end{equation}
(The expression for the regular part applies when $r \gg d$, the oscillatory part when $k_F r \gg 1$; both hold when $r > 2X_0$). For a uniform density $n_{wh}$ of whole charge defects, $ \Gamma_\phi^{wh, n} = \Gamma_{\phi>}^{wh, n} + \Gamma_{\phi<}^{wh, n}$, where $\Gamma_{\phi>}^{wh, n} = \frac{n_{wh} \alpha^4\tau}{2\hbar^2} \, X_0^2 \big(\frac{e^2 q_{TF}}{4\pi \epsilon_0 \epsilon_r}\big)^2 \, I_> $ and
\begin{equation}
\arraycolsep 0.3 ex
\begin{array}{rl} 
\displaystyle I_> = & \displaystyle
\int_{2}^\infty d\tilde{r} \, \tilde{r} \int_0^{2\pi} d\theta \bigg\{ \bigg[\frac{1 + q_{TF}d}{q_{TF}^3 X_0^3[(\tilde{r} \cos \theta + 1)^2 + \tilde{r}^2\sin^2\theta]^{3/2}} - e^{-2k_Fd} \frac{\sin2k_FX_0 \sqrt{[(\tilde{r} \cos \theta + 1)^2 + \tilde{r}^2\sin^2\theta]}}{(2k_FX_0)^2[(\tilde{r} \cos \theta + 1)^2 + \tilde{r}^2\sin^2\theta]}\bigg] \\ [3ex]
- & \displaystyle \bigg[\frac{1 + q_{TF}d}{q_{TF}^3 X_0^3[(\tilde{r} \cos \theta - 1)^2 + \tilde{r}^2\sin^2\theta]^{3/2}} - e^{-2k_Fd} \frac{\sin2k_FX_0\sqrt{[(\tilde{r} \cos \theta - 1)^2 + \tilde{r}^2\sin^2\theta]}}{(2k_FX_0)^2[(\tilde{r} \cos \theta - 1)^2 + \tilde{r}^2\sin^2\theta]}\bigg] \bigg\}^2.
\end{array}
\end{equation}
For $r < 2X_0$, letting ${\bm q} = 2k_F \tilde{\bm q}$, $\Gamma^{wh, n}_{\phi<} = \frac{n_{wh}\alpha^4k_F^2X_0^2\tau}{8 \pi^4\hbar^2} \, (\frac{e^2}{2 \epsilon_0 \epsilon_r})^2 I_< $
\begin{equation}\label{Gammanwh}
\arraycolsep 0.3 ex
\begin{array}{rl}
\displaystyle I_< = & \displaystyle \int_0^2 d\tilde{r} \, \tilde{r} \int_0^{2\pi} d\theta \int_0^\infty d \tilde{q} \, \tilde{q} \int_0^{2\pi} d\phi \int_0^\infty d \tilde{q}' \tilde{q}' \int_0^{2\pi} d\phi' \\ [3ex] 
& \displaystyle (\cos 2 k_F\tilde{\bm q} \cdot {\bm R}_{L\perp} - \cos 2 k_F\tilde{\bm q} \cdot {\bm R}_{R\perp}) \, e^{-2k_Fd\tilde{q}} \bigg[\frac{\Theta(1 - \tilde{q})}{\tilde{q} + \tilde{q}_{TF}} + \frac{\Theta(\tilde{q} - 1)}{\tilde{q} + \tilde{q}_{TF} \, \big\{ 1 - \big[ 1 - \big( \frac{1}{\tilde{q}} \big)^2 \big]^{1/2}  \big\} } \bigg] 
\\ [3ex]
& \displaystyle (\cos 2 k_F\tilde{{\bm q}}' \cdot {\bm R}_{L\perp} - \cos 2 k_F\tilde{\bm q}' \cdot {\bm R}_{R\perp}) \, e^{-2k_F d\tilde{q}'} \bigg[\frac{\Theta(1 - \tilde{q}')}{\tilde{q}' + \tilde{q}_{TF}} + \frac{\Theta(\tilde{q}' - 1)}{\tilde{q}' + \tilde{q}_{TF} \, \big\{ 1 - \big[ 1 - \big( \frac{1}{\tilde{q}'} \big)^2 \big]^{1/2}  \big\} } \bigg] .
\end{array}
\end{equation}
The vector $\tilde{\bm q} = \tilde{q} \, (\cos\phi, \sin\phi)$, this \textit{defines} the polar angle $\phi$ of ${\bm q}$, and 
\begin{equation}
\arraycolsep 0.3 ex
\begin{array}{rl}
\displaystyle \tilde{\bm q} \cdot {\bm R}_{R\perp} = & \displaystyle \tilde{q} \, X_0 \, [\cos\phi \, (\tilde{r} \cos \theta - 1) + \sin\phi \, (\tilde{r} \sin\theta)] \\ [1ex]
\displaystyle \tilde{\bm q} \cdot {\bm R}_{L\perp} = & \displaystyle \tilde{q} \, X_0 \, [\cos\phi \, (\tilde{r} \cos \theta + 1) + \sin\phi \, (\tilde{r} \sin\theta)]. 
\end{array}
\end{equation}

\subsubsection{Dephasing rate for dipole charge defects}

Here $U_{dip}^D = 2 {\bm l}_\perp \cdot\pd{U_{scr}^D}{{\bm R}_{D\perp}} + 2 l_z \, \pd{U_{scr}^D}{d}$, and $ \Gamma_\phi^{dip, n} = \Gamma_{\phi>}^{dip, n} + \Gamma_{\phi<}^{dip, n}$. For $r > 2X_0$, 
$\Gamma_{\phi>}^{dip, n} = \frac{n_{dip} \alpha^4 \tau}{2\hbar^2 X_0^4} \, (\frac{e^2}{2\pi \epsilon_0 \epsilon_r q_{TF}^2})^2 J_>$ with
\begin{equation}
\arraycolsep 0.3 ex
\begin{array}{rl}
\displaystyle J_> = & \displaystyle \int_2^\infty d\tilde{r} \, \tilde{r} \int_0^{2\pi} d\theta \bigg\{ \bigg[ \frac{q_{TF} l_z}{[(\tilde{r} \cos \theta + 1)^2 + \tilde{r}^2\sin^2\theta]^{3/2}} - \frac{3 \, (1 + q_{TF}d ) (\hat {\bm R}_{L\perp} \cdot{\bm l}_\perp/X_0)}{[(\tilde{r} \cos \theta + 1)^2 + \tilde{r}^2\sin^2\theta]^2} \bigg] \\ [3ex]
- & \displaystyle \bigg[ \frac{q_{TF} l_z}{[(\tilde{r} \cos \theta - 1)^2 + \tilde{r}^2\sin^2\theta]^{3/2}} - \frac{3 \, (1 + q_{TF}d ) (\hat {\bm R}_{R\perp} \cdot{\bm l}_\perp/X_0)}{[(\tilde{r} \cos \theta - 1)^2 + \tilde{r}^2\sin^2\theta]^2} \bigg]\bigg\}^2.
\end{array}
\end{equation}
For a uniform density $n_{dip}$ and ${\bm l} = l \, \hat{\bm l}$, we have $\Gamma^{dip, n}_{\phi<} = \frac{n_{dip} \alpha^4 X_0^2l^2 \tau}{2\hbar^2} \, \big(\frac{e^2}{\epsilon_0 \epsilon_r}\big)^2 \big(\frac{k_F}{\pi}\big)^4 J_<$
\begin{equation}\label{Gammandip}
\arraycolsep 0.3 ex
\begin{array}{rl}
\displaystyle J_< = & \displaystyle \int_0^2 d\tilde{r} \, \tilde{r} \int_0^{2\pi} d\theta \int_0^\infty d^2 \tilde{q}\, [(\hat{\bm l}_\perp \cdot \tilde{\bm q} \sin 2k_F\tilde{\bm q} \cdot {\bm R}_{L\perp} + \hat{l}_z \, \tilde{q} \, \cos 2k_F\tilde{\bm q} \cdot {\bm R}_{L\perp}) - (\hat{\bm l}_\perp \cdot \tilde{\bm q} \sin 2k_F \tilde{\bm q} \cdot {\bm R}_{R\perp} + \hat{l}_z \, \tilde{q} \, \cos 2k_F \tilde{\bm q} \cdot {\bm R}_{R\perp})] \\ [3ex]

& \displaystyle e^{-2k_Fd\tilde{q}} \bigg[\frac{\Theta(1 - \tilde{q})}{\tilde{q} + \tilde{q}_{TF}} + \frac{\Theta(\tilde{q} - 1)}{\tilde{q} + \tilde{q}_{TF} \, \big\{ 1 - \big[ 1 - \big( \frac{1}{\tilde{q}} \big)^2 \big]^{1/2}  \big\} } \bigg]  \\ [3ex]

& \displaystyle \int_0^\infty d^2 \tilde{q}' \, [(\hat{\bm l}_\perp \cdot \tilde{\bm q}' \sin2k_F\tilde{\bm q}' \cdot {\bm R}_{L\perp} + \hat{l}_z \, \tilde{q}' \, \cos 2k_F\tilde{\bm q}' \cdot {\bm R}_{L\perp}) - (\hat{\bm l}_\perp \cdot \tilde{\bm q}' \sin2k_F\tilde{\bm q}' \cdot {\bm R}_{R\perp} + \hat{l}_z \, \tilde{q}' \, \cos2k_F\tilde{\bm q}' \cdot {\bm R}_{R\perp})] \\ [3ex]

& \displaystyle e^{-2k_Fd\tilde{q}'} \bigg[\frac{\Theta(1 - \tilde{q}')}{\tilde{q}' + \tilde{q}_{TF}} + \frac{\Theta(\tilde{q}' - 1)}{\tilde{q}' + \tilde{q}_{TF} \, \big\{ 1 - \big[ 1 - \big( \frac{1}{\tilde{q}'} \big)^2 \big]^{1/2}  \big\} } \bigg].
\end{array}
\end{equation}

\end{widetext}

\end{document}